# Evidence of local superconductivity in granular Bi nanowires fabricated by electrodeposition

Mingliang Tian, Nitesh Kumar, Thomas E. Mallouk and Moses H.W. Chan
Center for Nanoscale Science and Department of Physics, 104 Davey Laboratory, the Pennsylvania State University, University Park, Pennsylvania 16802

Abstract

An unusual enhancement of resistance (i.e., a "superresistivity") below a certain characteristic temperature $T_{sr}$ was observed in granular Bi nanowires. This "superresistive" state was found to be dependent on the applied magnetic field (H) as well as the excitation current (I). The suppression of $T_{sr}$ by magnetic field resembles that of a superconductor. The observed superresistivity appears to be related to the nucleation of local superconductivity inside the granular nanowire without long-range phase coherence. The phenomenon is reminiscent of the "Bose-insulator" observed previously in ultra thin two-dimensional (2D) superconducting films and 3D percolative superconducting films.

Bulk bismuth (Bi) has a rhombohedral crystal structure and displays semimetal properties down to at least 50 mK without showing evidence of superconductivity [1,2]. Recently, we reported superconductivity in Bi nanowires with $T_c$ of 7.2 K and 8.3 K [3]. These nanowires, fabricated by electrodepositing Bi into porous polycarbonate membranes, showed granular morphology consisting of crystalline rhombohedral Bi grains of a few nanometers to 15 nm. However, in spite of our effort in following similar fabrication conditions and protocols, superconductivity with a sharp resistance drop at $T_c$ was found in only 18 out of a total of 38 samples studied. In contrast, the other 20 granular nanowire samples showed non-superconducting behavior down to 0.47 K. Transmission electron microscopy (TEM) and electron diffraction (ED) measurements showed interesting contrast in the morphology of the superconducting and the non-superconducting granular nanowires. The rhombohedral grains in the superconducting wires are aligned along the [001] direction within an angular distribution of 19° and the wire morphology is uniform along the wire and from wire to wire [3]. In contrast, the grains in the non-superconducting nanowires showed random orientations [3]. Because the transition temperatures of 7.2 K and 8.3 K in superconducting wires are identical to those of the tetragonal high pressure phases Bi-III [4,5,6,7] and body-centered cubic Bi-V [4,5,8,9,10], we suggested that the grain boundaries between the grains with [001] orientation may assume the same crystal structures as those of Bi under high pressure due to structural reconstruction or local distortion. The [001] alignment of the grains may allow a superconducting path along the atomically thin high-pressure phase percolating through at least one or a few granular wires in an array of wires embedded in the porous membrane. The high pressure phases were not detected by X-ray or TEM since the thickness of boundary layer is only on the order of a few atomic layers.

In this paper, we report a systematic study of the non-superconducting granular Bi nanowires. In the 20 non-superconducting granular wires we investigated, nine of them showed an unusual "superresistive" behavior, specifically, an abrupt enhancement of resistance (R) below a well-defined temperature, $T_{sr}$. The value of $T_{sr}$ depends on the details of the sample and varies from sample to sample. The possible origins will be discussed below. The transport properties in the samples with "superresistivity" below $T_{sr}$ are found to depend not only on the applied magnetic field (H) but also the applied excitation current (I). By increasing H, $T_{sr}$ is suppressed correspondingly and a phase boundary of the superresistive state can be mapped out. For H exceeding a critical value, $H_{sr}$, the enhanced resistance can be completely suppressed resulting in a smooth semiconductor-like R-T curve from ~ 60 K down to 0.47 K. The $H_{sr}$-T phase boundary of the superresistive state resembles that of a superconductor. The observed superresistivity appears to be related to the nucleation of 'local' superconductivity without long-range phase coherence. The phenomenon is reminiscent of that reported in ultrathin two-dimensional (2D) granular Sn [11], Al[12], In, Ga and Pb films [13,14] and three-dimensional (3D) granular Al [15], Al-Ge mixture films [16] on the insulating side of the superconductor-insulator transition (SIT), as well as in percolative superconducting Pb films below the percolation threshold [17].

Granular Bi nanowires used in this work were made by electrochemically depositing Bi into commercially available porous polycarbonate (PC) membranes at room temperature (the details of the fabrication process have been described in ref. 3). We found that the granular nanowires can be achieved with a deposition voltage between -2.0 V and -3.5 V. The diameter and length of the nanowires are controlled by the pore size and the thickness of the membrane [3,18,19] (The actual diameter of the resulting nanowires is usually larger than the quoted pore size by manufacturers, the possible origins have been discussed in literatures [18,20]). In this report, all Bi nanowires have an actual diameter of 70 nm and a length ($L$) of 6 $\mu$m. Freestanding nanowires were obtained by dissolving the PC membrane in dichloromethane for TEM imaging.

Transport measurements were carried out on Bi nanowire arrays embedded in the PC membrane with a Physical Properties Measurement System

(PPMS), equipped with a $^3$He insert and a superconducting magnet. Details about the transport measurements are reported earlier. [21, 22] The total resistance, R, of the system with this configuration consists of the contributions from the Bi nanowires, the metallic electrodes and the point contacts between the electrodes and Bi nanowires. While the resistance of the metallic electrodes (i.e., Ag) is negligibly small (on the order of < 0.1 Ω), the point contact resistance of diameter (*d*) might not be negligible and can be estimated approximately by means of the Sharvin formula [23], $R_{sh} = 4\rho_0 \ell_e / 3\pi d^2$, where $\rho_0$ is the resistivity of bulk Bi and $\ell_e$ the electron mean-free path. Using $\rho_0 \ell_e \sim 10^{-8} \Omega.cm^2$ [24], $R_{sh}$ is eatimated to be 80 Ω for 70 nm Bi nanowires. This value is added to the total resistance as a series component to each Bi nanowire. If we make the assumption that the wires in each array are identical, the measured total resistance can be expressed as $R = (R_{nw} + 2R_{sh})/N$, where $R_{nw}$ is the resistance of each individual Bi nanowire in the array and *N* is the total number of the nanowires making contact with the electrodes. For a nanowire with *L*= 6 μm and *d* = *70 nm*, and taking a room temperature resistivity value, $\rho$ = 315 ~ 850 μΩ.cm[25], we get $R_{nw} = 4\rho L/\pi d^2$ to be on the order of 4.9 ~ 13.3 *k*Ω at room temperature. Since $R_{nw} \gg R_{sh}$, the measured total resistance is dominated by the Bi nanowires.

Figure 1 show R vs. T curves measured at zero magnetic field for the nine granular Bi nanowire samples showing the superresistive behavior, Each sample is consisted of an array of non-intersecting 30 to 60 wires making parallel electrical contact to the electrodes. The value of the resistances are normalized to that at 100 K. All samples showed an anomaly at $T_{sr}$. Below $T_{sr}$, the resistance either increases sharply or displays a small drop first and then increases with decreasing temperature. The onset value of $T_{sr}$ varies from sample to sample, and four samples showed a $T_{sr}$ at 7.1 ± 0.2 K or 8.1± 0.2 K, two samples at 3.6 ± 0.2 K and three samples at 5.8 ± 0.2 K. It is interesting that each sample shows a sharp anomaly at a single $T_{sr}$, and the resistance at low temperature shows a significant offset from the value extrapolated from T >$T_{sr}$. It appears at least a very significant fraction of the nanowires in each of the nine samples exhibit the same resistance anomaly at the same $T_{sr}$. This means the value of $T_{sr}$ depends sensitively on the exact growth conditions during the electrodeposition procedure, which we have not been able to completely manipulate or control them from sample to sample. Other than different $T_{sr}$ among different samples, all nine Bi nanowire samples show similar dependence on temperature, magnetic field and excitation current. Here we focused our attention on one of these samples, namely sample #B1 with $T_{sr}$ = 5.8 ± 0.2 K. The typical TEM image of the sample #B1 is shown in the inset of Fig. 2 (a). This sample was deposited at a potential of -2.45V and the nanowires show granular morphology with grain size ranging from a few nanometers up to 40 nm.

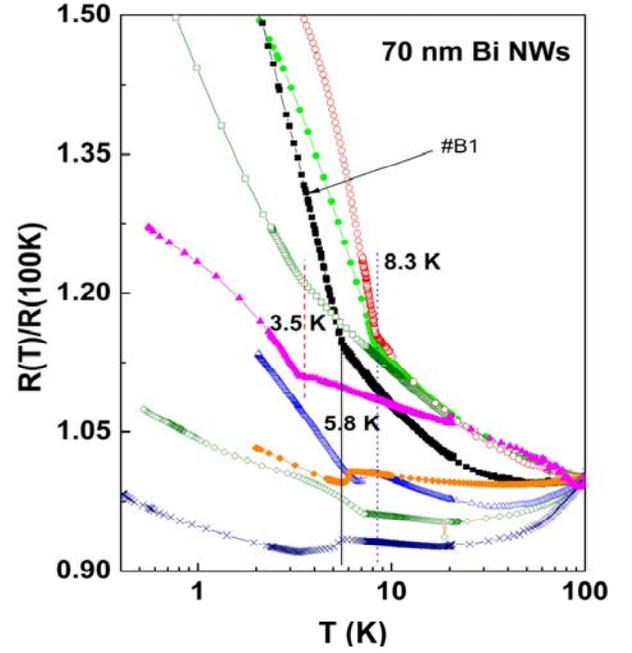

FIG.1. R vs. T curves of nine granular Bi nanowire samples (d=70 nm, L=6 μ m) showing a "superresistive" behavior at a well defined $T_{sr}$. Sample #B1 shows a $T_{sr}$ of 5.8 K, which will be discussed specifically in context.

Fig. 2 (a) shows the R-T curves of sample #B1 with an excitation current of 50 nA under perpendicular magnetic fields H of 0, 3.0, 4.0, 5.0 and 8.0 *k*Oe. Since the measured total resistance R= 248 Ω at 300 K, the number *N* of the wires in this array is estimated to be on the order of ~ 20 - 54 using $\rho$ = 315 ~ 850 μΩ.cm [25]. The R-T curves

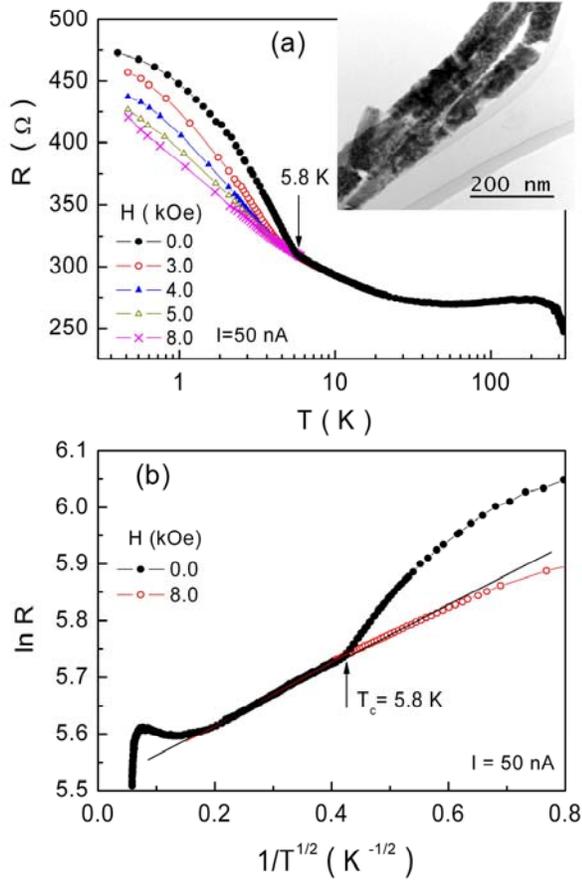

FIG. 2 (a) R vs. T curves of granular Bi nanowire sample #B1 under different perpendicular magnetic fields H, measured with a small dc excitation current of 50 nA. The inset shows a TEM image of the wires. (b) lnR v.s. $1/T^{1/2}$ curves at two specific magnetic fields of H=0 $k$Oe and 8.0 $k$Oe.

show a broad maximum near 200 K and insulating behavior below 60 K. Such a broad maximum in R is a common feature of semimetallic Bi nanowires [25, 26, 27] and can be understood as a consequence of the competition between the temperature dependence of the carrier concentration ($n$) and the carrier mobility ($\mu_0$) in determining the resistivity $\rho = 1/ne\mu_0$. While n decreases, $\mu_0$ increases with decreasing temperature. Since n is nearly constant below 100 K, a metallic behavior or a saturation in resistance is expected at low temperatures. However, the situation in Bi nanowires is more complicated due to the finite-size effect [26], weak-localization [28] and enhanced e-e interactions [26, 29] or other mechanisms [30, 31].

In sample #B1, we found that, under a magnetic field of H = 8.0 $k$Oe, the R-T curve shows a smooth insulating behavior from 60 K down to 0.42 K, and its resistance $lnR$ scales linearly with $T^{-1/2}$ between 2 K and 55 K, as shown in Fig. 2(b). This data can be expressed as $R = R_0 \exp[T_0/T]^{1/2}$ with fitting parameters of $R_0$ = 245 $\Omega$ and $T_0$ = 0.31 K. The exponential divergence in resistance is consistent with the model of strong localization with variable-range hopping (VRH) for a finite 1D wire [32] or the Coulomb gap model of Efros-Shklovskii (ES) [33]. The ES model is valid in both the 2D and 3D strong localization limits due to the Coulomb interaction. According to the VRH or ES model, the granular Bi sample #B1 under a magnetic field of H ≥ 8.0 $k$Oe is a "true insulator" below 60 K with a small activation energy $T_0$ ~ 0.31 K.

At H = 5.0 $k$Oe and below, the measured resistance show an abrupt enhancement below $T_{sr}$ that rides on top of the smooth R-T curve obtained at H = 8.0 $k$Oe. The onset, $T_{sr}$, of the "superresistive" behavior increases with decreasing H, and it reaches 5.8 ± 0.2 K at H = 0 Oe. Fig. 3 (a) shows R versus H curves at different temperatures. For T > 6.0 K, a small positive magnetoresistance (MR) is found. Below 5.5 K, R-H curves show a plateau in the low field region, and then a negative MR in higher field until a critical value, $H_{sr}$ before exhibiting the same positive MR behavior as observed for T > 6.0 K. The value of $H_{sr}$ decreases with temperature and extrapolates to zero near 5.8 K. The overall feature of the R-T and R-H curves below 5.8 K is similar to that reported for a percolating superconducting Pb film below the percolation threshold [17] and in granular Al-Ge mixture films on the insulating side of the SIT [16]. These percolating granular films showed a non-superconducting behavior and a sharp resistance enhancement below the bulk $T_c$ of Pb or Al at zero magnetic field, which is similar to that shown in Figs.1-3.

Similar "superresistive" behavior was also observed in ultrathin 2D granular Sn[11], Al [12], In, Ga and Pb films [13, 14] when the normal state sheet resistance of the films is tuned to be on the order of the quantum resistance, $R_Q = h/4e^2 = 6.4$ $k\Omega$. The rapid increase of electrical resistance below the

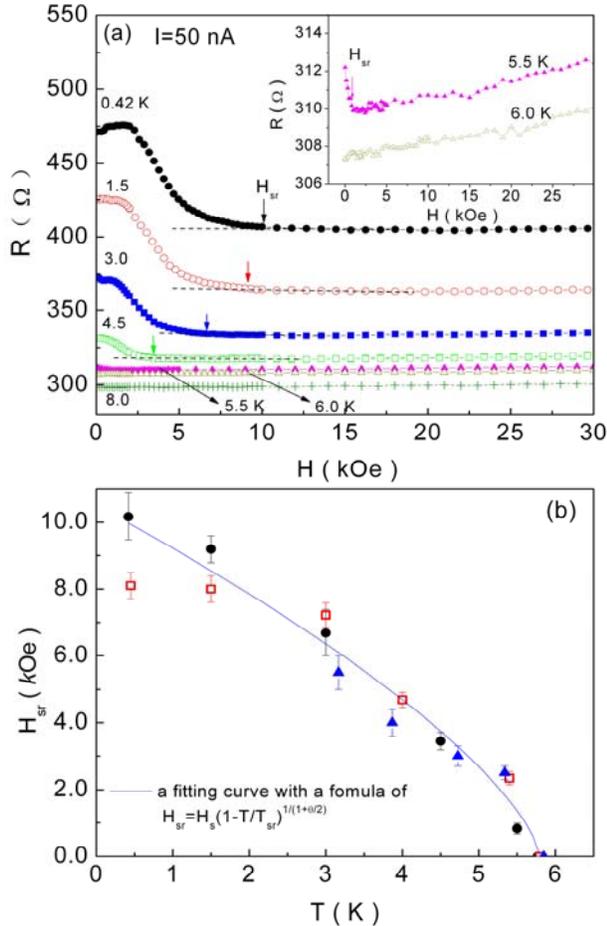

FIG. 3 (a) R vs. H curves of granular Bi nanowire sample #B1, measured at different temperatures with a small dc excitation current of 50 nA. The inset is a blow-up of the R-H curves at 5.5 K and 6.0 K, respectively. (b) the phase diagram of $H_{sr}$ vs. T, obtained from the R-H and R-T curves shown in Figs. 3(a), 4(b) and .5.

bulk $T_c$ of granular superconducting films was interpreted as the consequence of the localized pairing of electrons without global phase coherence of the superconducting order parameter. In other words, the "superresistive" insulating state in these ultrathin granular films is not a true insulator, and the Cooper pairs are postulated to survive inside each individual grain. As a result, the conductivity is dominated by single-electron tunneling processes between the neighboring superconducting islands. The reduction of the gap of these superconducting islands by the application of a magnetic field results in a reduction of the resistance of the films, i.e., a negative MR effect. The similarity in the transport behavior of our granular Bi nanowires suggests that the observed "superresistivity" below 5.8 K may have the same physical origin. In contrast to granular Sn, In, Pb or Al films where the bulk materials are well-known superconductors, bulk Bi is not a superconductor and it is natural to wonder about the origin of the local superconductivity in granular Bi nanowires.

As noted above, superconductivity was found in granular Bi nanowires when the rhombohedral grains in these wires are oriented along the [001] direction [3]. We have suggested that the superconductivity takes place along the boundaries between the aligned grains. It is possible that due to surface reconstruction, a few atomic layers of atoms at the boundaries between the Bi grains may take on the structure of the superconducting high pressure Bi-III and Bi-V phases. This interpretation is supported by the observation that the onset of superconductivity in these oriented granular wires takes place at 8.3 K and 7.2 K, the superconducting transition temperatures of these high-pressure phases. It is reasonable to expect that the boundaries between grains that are randomly oriented are not superconducting. However, one cannot exclude the possibility that some neighboring grains have local [001] orientation, which leads to local superconductivity at these local boundaries without a global long-range phase coherence. As seen in Fig.1, the samples showing an anomaly near 3.6 K, 7.2 or 8.3 K may correspond to the case of local superconductivity due to the formation of high pressure phases Bi-II, III and V, respectively. What is then the origin of the $T_c$ of 5.8 ± 0.2 K as seen in sample #B1? It is known that thin amorphous Bi film has a $T_c$ near 5.8 ~ 6.0 K [34] at zero magnetic field. The observed "superresistivity" in #B1 near 5.8 ± 0.2 K is very likely related to the nucleation of an amorphous Bi phase residing at grain boundaries. Since the majority of the crystalline rhombohedral Bi grains and the boundaries are still non-superconducting, these amorphous islands do not exhibit long-range phase coherence.

The zero field R-T curve in Fig. 2 and the R-H curves in Fig.3 (a) measured at different temperatures showed that for a certain T, superresistive behavior is found below a specific magnetic field $H_{sr}$, defined from the point when the resistance deviates from the linear baseline ~ 0.7

Ω. These data points, defining a $H_{sr}$-T phase diagram, are shown in Fig. 3 (b) as filled solid circles "●" (the scattering of the R data as shown in the inset of Fig. 3 (a) is less than 0.5 Ω and the error bar of the $H_{sr}$ is determined from the variation of H when the R deviates from the baseline ~ 1.4 Ω). Data points shown as open squares and solid triangles are deduced from different measurements and will be discussed below. The $H_{sr}$ -T phase diagram resembles that of a 2D percolation superconducting film, where the temperature dependence of the upper critical magnetic field $H_{c2}$ is predicted to follow the relation $H_{c2}(T) \propto \xi_s^{-2} \propto (T_c - T)^{1/(1+\theta/2)}$ with θ ~ 0.9 when the percolation correlation length $\xi_p$ becomes longer than the effective superconducting phase coherence length $\xi_s$ [17, 35]. The solid line shown in Fig. 3(b) is a fit of the six data points shown as the filled solid circles according to the same formula, namely, $H_{sr}(T) = H_s(1 - T/T_c)^{1/(1+\theta/2)}$ with θ ~ 0.9, $T_c$ = 5.8 K and a fitting parameter $H_s$ =10.50 ± 0.32 kOe. The quality of the fit suggests that $H_{sr}$ can be attributed to the upper critical field $H_{c2}$ of a percolative superconductor. Above $H_{sr}$, the superconducting gap in each localized superconducting Bi island is completely destroyed and this leads to a smooth insulating R - T curve, consistent with a variable-range hopping process as shown in Fig. 2 (b).

Fig. 4 (a) shows the R-T curves measured at different dc excitation currents. It is known that the mechanism for superconducting to insulating transition (SIT) in a granular system is the competition between the intergrain charging energy ($E_c$) and the Josephson coupling energy ($E_J$) between neighboring grains. When $E_c < E_j$, the films are superconducting. In the opposite limit of $E_c > E_j$, the electrostatic energy suppresses charge fluctuations and localizes Cooper pairs inside each grain, and thus leads to an insulating behavior. The resistance close to the SIT shows a drop at the bulk $T_c$, and then develops an upturn into the insulating state at lower temperatures. This mechanism has been successfully used to explain not only the behavior of 2D granular films[11, 13, 14] but also for Josephson junctions [36]/junction arrays [37] and 3D granular system [16] (note: the localized pairing mechanism might be also appropriate for the electrically insulating behavior in homogeneous films based on the recent experiment on homogeneous Bi film by Stewart *et al.* [38], but the detailed picture for the local pairing in a homogeneous film remains unclear). By applying a dc bias onto a granular insulating system with local superconductivity, one expects that the charging energy might be, to some extent, suppressed and hence a current-induced insulator to superconductor transition may appear due to the recovery of Josephson coupling. Such a current-induced effect is seen in Fig. 4 (a). With an excitation current of I ≤ 5.0 μA, the R-T curves showed an upward kink at 5.8 K and a rapid increase in resistance below 5.8 K. At an excitation current of 10 μA, a tiny resistance drop is found at 5.8 K prior to the appearance of the insulating behavior. The magnitude of the drop at 5.8 K increases with increasing dc bias (inset of Fig. 4(a)). When I = 25 $\mu$A, the resistance drop from 5.8 K to 0.42 K reaches about 0.7 % of the total normal state resistance $R_N$ (i.e., $\Delta R/R_N \sim 0.7\%$ ), and the R-T curve becomes nearly temperature independent below 5.8 K. This dc current-induced insulator to metal-like transition is not an artifact of the heating effect because the kink position at 5.8 K does not shift with the excitation current, as shown in the inset of Fig. 4 (a). Fig. 4 (b) show the R-T curves, measured with an excitation current of 25 $\mu$A under different perpendicular magnetic-fields. The temperature, $T_{sr}$, at the kink decreases with increasing applied H and finally becomes unresolvable at H ≥ 5.5 kOe. This behavior resembles to that seen in a superconductor, where the critical temperature $T_c$ is suppressed by an applied magnetic field. The phase boundary deduced from the kink position in Fig. 4 (b), shown in Fig. 3 (b) as solid triangles "▲", is consistent with that of the same curve defined from Fig. 3 (a). We thus attribute the current-induced tiny resistance drop at 5.8 K to the partial recovery of superconducting coupling between the localized superconducting islands. A similar current-induced insulator to superconductor transition was also reported by Wu et al in an ultra thin granular Al film [12]. They found that the granular Al film showed an insulating behavior when measured at a low excitation current of I < 60 pA, but was superconducting at I > 60 nA. The resistance drop reaches about 90 % of the normal state resistance from $T_c$ down to 0.4 K, which is much larger than that observed in granular Bi sample #B1. The current-induced small resistance drop at $T_{sr}$ in granular Bi nanowires provides an indication that the number of superconducting islands in #B1 is likely very small and therefore the Josephson-coupling between these islands is extremely weak even under a large dc bias.

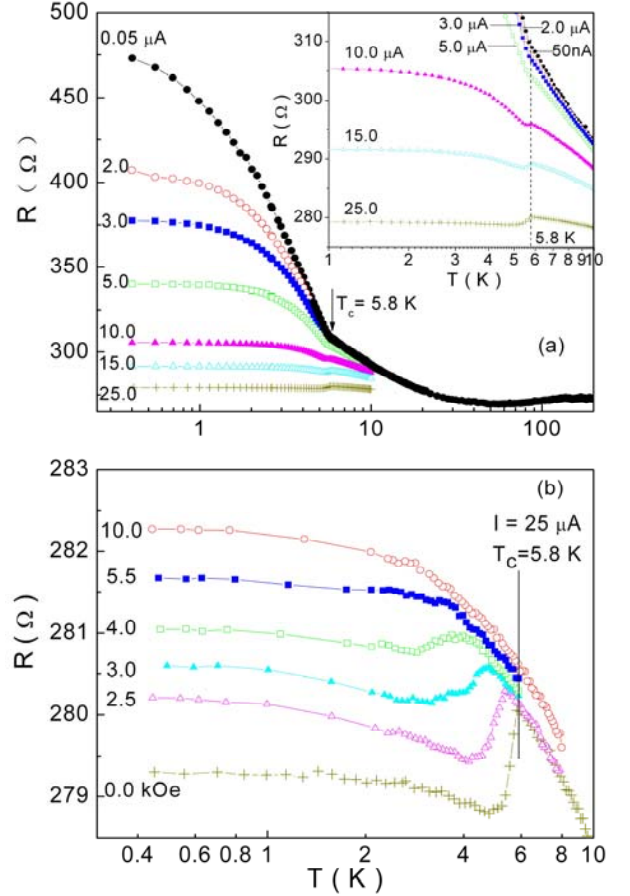

FIG.4. R vs. T of sample #B1, measured with different dc excitation currents. The inset is a blow-up near the $T_c$. (b) R vs. T curves under different H, measured with a higher dc current of 25 μA.

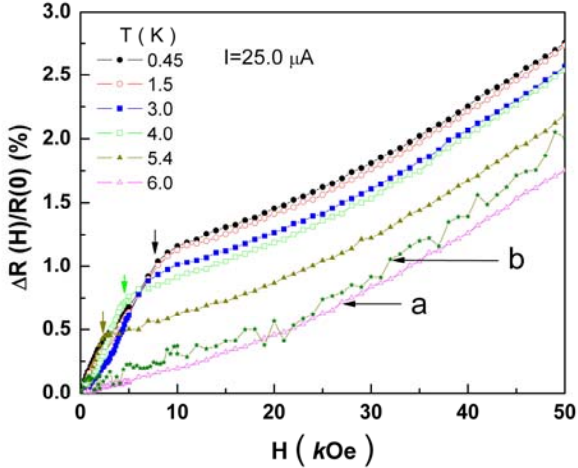

FIG. 5 Magnetoresistance $\Delta R/R(0)$ vs. H of sample #B1 at different T, measured with a higher dc current, 25 µA. A positive MR effect was seen, which is opposite to that shown in Fig. 3 (a) measured with a small excitation current of 50 nA.

We note that the total reduction of the wire resistance under an excitation current is unlikely to be due solely to the enhanced Josephson coupling between the superconducting islands. This is the case because the current dependent effect of the resistance was seen not only below $T_{sr}$ but also above $T_{sr}$ (as shown in Fig.4a). Since the resistance above $T_{sr}$ is due to single-electron hopping processes[32] between the strong localized states, the nonlinear dependence of the resistance on the excitation energy is expected, according to the VRH theory, when the excitation energy is larger than $k_B T$ (i.e., $dI/dV \sim \exp(V/k_B T)$ [32]). Hence, the substantial decrease of the total resistance below $T_{sr}$ by increasing the current from 50 nA to 25 $\mu$A originates from both the VRH mechanism and the partial recovery of Josephson coupling between the localized pairing islands. A clear signature of the latter process is the tiny drop of the resistance at 5.8 ± 0.2 K under a large excitation current.

Fig. 5 shows curves of $\Delta R/R(0)$ vs. H , measured with a higher excitation current of 25 $\mu$A at different temperatures. At T = 6.0 K, R increases smoothly with H and this positive MR curve (labeled "a") is similar to that measured with a small excitation current of 50 nA at the same temperature, as shown in curve "b" (i.e., the same curve as shown in Fig. 3 (a) at 6.0 K). Both curves appear to collapse onto each other and thus indicate that the MR effect at T > 6.0 K is independent of the excitation current. We attribute this positive MR at T > 6.0 K to the normal galvanomagnetic MR due to the Lorenz force on electrons. It is worth noting that, at T < 5.8 K, the $\Delta R/R(0)$ -H curves measured at 25 $\mu$A showed an opposite behavior compared with those shown in Fig. 3 (a). The MR, instead of being negative as in Fig. 3, is positive and the curves show a convex shape in the low field region with a shoulder at a characteristic value indicated by the arrows. When the temperature increases, the shoulder position shifts to lower values of the magnetic field. If we define the magnetic field at the shoulder position as $H_{sr}$ and then add the $H_{sr}$ data into Fig. 3 (b) as open squares "□", we find that the phase boundary defined from Fig. 5 is in reasonable agreement with those defined from Fig. 3 (a) and Fig. 4 (b) within an uncertainty of 25%. These data indicate that the resistance shoulder in $\Delta R/R(0)$ -H curves measured at 25 $\mu$A probably corresponds to the upper critical field $H_{c2}$ of the local superconducting islands.

In addition to the 18 samples showing superconductivity and 9 samples showing superresistivity, 11 samples showed a smooth insulating behavior down to 0.47 K. Since our measurements were performed on nanowire arrays, the experiments do not allow us to quantitatively determine whether the observed smooth insulating behavior is due to the absence of local pairs. If a small number of local pairs survive only in some of the granular wires in the array, they may not show up in resistance measurement under a parallel configuration of nanowires. Even if in sample #B1, it is still difficult to estimate how exact percentage of the boundaries show superconductivity with local pairs, but our data clearly indicate that the transport behavior of granular Bi nanowires strongly depends on the exact local morphology of an individual nanowire and the structural orientations between grains. They can be superconducting, superresistive or insulating due to the formation of superconducting islands at grain boundaries. In fact, experimental magnetization evidence of superconductivity in Bi bicrystals with a large-angle (>30°) twisting type crystallite interface were reported by Muntyanu et al.[39], two unknown superconducting phases with $T_c$ ~ 8.4 K and ~ 4.3 K were observed at the twisting interface. Very recently, Ye et al.[40] reported an evidence of superconductivity at $T_c$ ~ 0.64 K in a rhombohedral single-crystal Bi nanowires, they speculated that this superconducting anomaly might be related to the surface of the nanowire. These data indicated that the transport properties (e.g., the $T_c$) of Bi are exceptionally sensitive to their microstructures and geometric size, which require further investigation through numerical calculations in terms of some specific configurations (e.g., boundary type or orientation and so on).

In conclusion, we report an unusual superresistive phenomenon in granular Bi nanowires that appears to be due to the nucleation of 'local' superconductivity without global phase coherence. The phenomenon is reminiscent of the those observed previously in ultra thin two-dimensional (2D) superconducting films and 3D percolative superconducting films.

**Acknowledgments**: We acknowledge helpful discussions with J. Jain and Y. Liu. This work was supported by the Center for Nanoscale Science (Penn State MRSEC) funded by NSF under Grant No. DMR-0213623.

[1] Kurti, F. E. Simon, Proc. R. Soc. London Ser. A **151**, 610 (1935).

[2] F.Y.Yang, K. Liu, K. Hong, D. H. Reich, P. C. Searson, C. L. Chien, Science **284**, 1335 (1999).

[3] M. L. Tian, J. G. Wang, N. Kumar, T. H. Han, Y. Kobayashi, T. E. Mallouk and M. H. W. Chan, Nano Lett. **6**, 2773 (2006).

[4] O. Degtyareva, M. I. McMahon, R. J. Nelmes, High Pressure Res., **24**, 319 (2004).

[5] H. Imasaki, J. H. Chen, T. Kikegawa, Rev. Sci. Instrum. **66**, 1388 (1995).

[6] B. Brandt, N. I. Ginzburg, Sov. Phys. JETP **12**, 1082 (1961); and ibid. **17**, 326 (1963).

[7] H. D. Stromberg, D. R. Stephens, J. Phys. Chem. Solids **25**, 1015 (1964).

[8] T. T. Chen, J. T. Chen, J. D. Leslie, H. J. T Smith, Phys. Rev. Lett. **22**, 526 (1969).

[9] N. B. Brandt, N. I. Ginzburg, Comtemp. Phys. **10**, 355 (1969).

[10] J. Wittig, Z. Phys. **195**, 215 (1966).

[11] D. B. Haviland, H. M. Jaeger, B. G. Orr, and A. M. Goldman, Phys. Rev. B **40**, 719 (1989). B. G. Orr, H. M. Jaeger, and A. M. Goldman, Phys. Rev. B **32**, 7586 (1985).

[12] W. H. Wu, and P. W. Adams, Phys. Rev. B 50, 13065 (1994); Phys. Rev. Lett. **73**, 1412 (1994).

[13] R. P. Barber, Jr., L. M. Merchant, A. La Porta, and R. C. Dynes, Phys. Rev. B **49**, 3409 (1994).

[14] B. G. Orr, H. M. Jaeger, and A. M. Goldman, Phys. Rev. B **32**, 7586 (1985).

[15] M. Kunchur, Y. Z. Zhang, P. Lindenfeld, W. L. McLean, and J. S. Brooks, Phys. Rev. B **36**, 4062 (1987).

[16] A. Gerber, A. Milner, G. Deutscher, M. Karpovsky, and A. Gladkikh, Phys. Rev. Lett. **78**, 4277 (1997).

[17] A. Gerber and G. Deutscher, Phys. Rev. Lett. **63**, 1184 (1989); A. Kapitulnik, and G. Deutscher, Phys. Rev. Lett. **49**, 1444 (1982).

[18] M. L. Tian, J. G.Wang, J. Kurtz, T. E. Mallouk, M. H. W. Chan, Nano Lett. **3**, 919 (2003).

[19] J. G. Wang, M. L. Tian, N. Kumar and T. E. Mallouk, *Nano Lett*. **5**, 1247 (2005).

[20] Z. L. Xiao, C. Y. Han, U. Welp, H. H. Wang, W. K. Kwok, G. A. Willing, J. M. Hiller, R. E. Cook, D. J. Miller, G. W. Crabtree, Nano Lett. **2**, 1293 (2002).

[21] M. L. Tian, J. G. Wang, J. S. Kurtz, Y. Liu, M. H. W. Chan, T. S. Mayer, T. E. Mallouk, Phys. Rev. B **71**, 104521 (2005); M. L. Tian, J. G. Wang, J. Snyder, J. Kurtz, Y. Liu, P. Schiffer, T. E. Mallouk, M. H. W. Chan, Appl. Phys. Lett. **83**, 1620 (2003).

[22] M. L. Tian, N. Kumar, S. Y. Xu, J. G. Wang, J. S. Kurtz, M. H. W. Chan, Phys. Rev. Lett. **95**, 076802 (2005); M. L. Tian, N. Kumar, J. G. Wang, S. Y. Xu, J. S. Kurtz, M. H. W. Chan, Phys. Rev. B **74**, 014515 (2006).

[23] Y. V. Sharvin, Sov. Phys. JETP **21**, 655 (1965).

[24] T. E. Huber, K. Celestine, and M. J. Graf, Phys. Rev. B **67**, 245317 (2003).

[25] T. W. Cornelius, M. E. Toimil-Molares, R. Neumann, and S. Karim, J. Appl. Phys. **100**, 114307 (2006).

[26] Z. B. Zhang, X. Z. Sun, M. S. Dresselhaus, J. Y. Ying, and J. Heremans, Phys. Rev. B **61**, 4850 (2000).

[27] J. Heremans and C. M. Thrush, Phys. Rev. B **59**, 12579 (1999).

[28] J. Heremans, C. M. Thrush, Z. Zhang, X. Sun, M. S. Dresselhaus, J. Y. Ying, and D. T. Morelli, Phys. Rev. B **58**, R10091(1988).

[29] D. E. Beutler and N. Giodano, Phys. Rev. B **38**, 8 (1988).

[30] T. E. Huber, A. Nikolaeva, D. Gitsu, L. Konopko and M. J. Graf, Physica E **37**, 194 (2007).

[31] A. Nikolaeva, D. Gitsu, L. Konopko, M. J. Graf, and T. E. Huber, Phys. Rev. B **77**, 075332 (2008).

[32] P. A. Lee, Phys. Rev. Lett. **53**, 2042 (1984).

[33] A. L. Efros and B. I. Shklovskii, J. Phys. C **8**, L49 (1975).

[34] S. Chakravarty, S. Kivelson, G. T. Zimanyi, B. I. Halperin, Phys. Rev. B **35**, 7256 (1987).

[35] A. Gerber, and G. Deutscher, Phys. Rev. B **35**, 3214 (1987).

[36] J. S. Penttila, U. Parts, P. J. Hakonen, M. A. Paalanen, and E. B. Sonin, Phys. Rev. Lett. **82**, 1004 (1999).

[37] H. S. J. van der Zant, W. J. Elion, L. J. Geerligs, and J. E. Mooij, Phys. Rev. B **54**, 10081 (1996).

[38] M. D. Stewart Jr., A. Yin, J. M. Xu, J. M. Valles Jr., Science **318**, 1273 (2007).

[39] F. M. Muntyanu, A. Gilewski, K. Nenkov, J. Warchulska, and A. J. Zaleski, Phys. Rev. B **73**, 132507 (2006).

[40] Z. X. Ye, H. Zhang, H. D. Liu, W. H. Wu and Z. P. Luo, Nanotechnology **19**, 1 (2008).